\documentstyle[aps,epsf,preprint,graphicx,prl]{revtex}

\begin{document}
\draft

\def\beq{\begin{equation}}
\def\eeq{\end{equation}}
\def\beqn{\begin{eqnarray}}
\def\eeqn{\end{eqnarray}}
\def\souligne#1{$\underline{\smash{ \hbox{#1}}}$}

\def\g {\large \mbox{\boldmath $g$}}
\def\lg {\mbox {\large {$g$}}}
\def\la {\mbox {\large {$a$}}}
\def\G {\mbox{\boldmath $G$}} 
\def\M {\mbox{\boldmath $M$}}
\def\T {\mbox{\boldmath $T$}}
\def\S {\mbox{\boldmath $S$}}
\def\m {\mbox{\boldmath $m$}}  
\def\X {\mbox{\boldmath $X$}} 
\def\E {\mbox{\boldmath $E$}} 
\def\A {\mbox{\boldmath $A$}} 
\def\B {\mbox{\boldmath $B$}} 
\def\1 {\mbox{\boldmath $1$}} 
\def\sig {\mbox{ \large \boldmath $\sigma$}}
\def\bvareps {\mbox{ \large \boldmath $\varepsilon$}}
\def\bvarphi {\mbox{ \large \boldmath $\varphi$}}
\def\SIG {\mbox{\boldmath $\Sigma$}}  
\def\W {\mbox{\boldmath $W$}}
\def\V {\mbox{\boldmath $V$}}
\def\RHO {\mbox{\large \boldmath $\rho$}}
\def\R {\mbox{\boldmath $R$}}

\def\bH {\mbox{\boldmath $\mathcal H$}}

\def\bcalG{\pmb{\cal G}}  

\def\beq{\begin{equation}}
\def\eeq{\end{equation}}
\def\souligne#1{$\underline{\smash{ \hbox{#1}}}$}

\def\g {\mbox{\boldmath $g$}}
\def\lg {\large \mbox{$g$}}
\def\G {\mbox{\boldmath $G$}}  
\def\sig {\mbox{\boldmath $\sigma$}}
\def\SIG {\mbox{\boldmath $\Sigma$}}  
\def\W {\mbox{\boldmath $W$}}
\def\V {\mbox{\boldmath $V$}}
\def\RHO {\mbox{\boldmath $\rho$}}
\def\R {\mbox{\boldmath $R$}}

\def\ed{\end{document}}

\title{Coupled-Cluster Theory for Systems of Bosons in External Traps}

\author{Lorenz S. Cederbaum\footnote{E-mail: Lorenz.Cederbaum@urz.uni-heidelberg.de}, 
 Ofir E. Alon\footnote{E-mail: ofir@tc.pci.uni-heidelberg.de}
 and Alexej I. Streltsov\footnote{E-mail: alexej@tc.pci.uni-heidelberg.de}}

\address{
Theoretische Chemie, Physikalisch-Chemisches Institut, Universit\"at Heidelberg,\break 
Im Neuenheimer Feld 229, D-69120 Heidelberg, Germany}

\maketitle

\begin{abstract}

A coupled-cluster approach for systems of $N$ bosons in external traps is developed.
In the coupled-cluster approach the exact many-body wavefunction is obtained by 
applying an exponential operator $\exp\{T\}$ to the ground configuration $\left|\phi_0\right>$.
The natural ground configuration for bosons is, of course, when all reside in a single orbital.
Because of this simple structure of $\left|\phi_0\right>$, 
the appearance of excitation operators $T=\sum_{n=1}^N T_n$ for bosons is much simpler than for fermions.
We can treat very large numbers of bosons with coupled-cluster expansions.
In a substantial part of this work, we address the issue of
size consistency for bosons and enquire whether truncated coupled-cluster expansions are size consistent.
We show that, in contrast to the familiar situation for fermions for which 
coupled-cluster expansions are size consistent, 
for bosons the answer to this question {\it depends} on the choice of ground configuration.  
Utilizing the natural ground configuration, working equations for the truncated 
coupled-cluster with $T=T_1+T_2$, i.e., coupled-cluster singles doubles (CCSD) are explicitly derived.
Finally, an illustrative numerical example for a condensate with 
up to $N=10000$ bosons in an harmonic trap is provided and analyzed.
The results are highly promising.

\end{abstract}


\section{Introduction}

Following the experimental demonstrations of Bose-Einstein condensates 
in dilute gases \cite{Ketterle,Wieman}, 
the problem of many bosonic atoms interacting in a trap potential has 
attracted an accelerated interest by the scientific community, 
see \cite{review_Legget,Stringari_book} and references therein.
There are many phenomena trapped bosons exhibit
that can be described quite well by the standard mean-field approach,
namely Gross-Pitaevskii (GP) theory \cite{Gross_Pitaevskii}, 
see \cite{review_Legget,Stringari_book} and reference therein.
Side-by-side, the necessity to go beyond mean-field
and describe many-body facets of trapped bosons
has become well-accepted and perused by the community,
see the review \cite{Minguzzi} and references therein.

The many-boson problem is difficult to tackle.
Consider, for instance, the standard configuration-interaction (CI) approach which employs a basis set expansion.
When the interaction between the $N$ bosons is substantial and/or many of them are present,
the number of configurations necessary to properly describe the correlated 
wavefunction quickly increases beyond computational reach and truncations become a must.
When truncations of the CI expansion are made, 
there are hints and evidences to slow convergence of the CI expansion, see, e.g., \cite{Haugest,CCI}. 
Evidently, development of other many-body methods which truncate the full configuration 
space in a different manner are of high relevance and actuality. 

Coupled-cluster theory was first formulated in nuclear physics 
by Coester \cite{CC_r1} and Coester and K\"ummel \cite{CC_r2}, 
and soon after was introduced to electron-structure theory by 
\v{C}i\v{z}ek \cite{CC_r3} and \v{C}i\v{z}ek and Paldus \cite{CC_r4}.
Coupled-cluster theory has since proven to be a very valuable and 
accurate approach in the many-fermion problem, see \cite{rev1,rev2,rev3} and references therein.
For atomic and molecular systems, coupled-cluster theory is currently considered to be 
one of the if not the most 
powerful many-body tool for calculating electron-correlation energies \cite{rev1,rev2,rev3},
also in relativistic systems \cite{Kaldor1_2}. 
In the coupled-cluster approach the exact many-body wavefunction is obtained by 
applying an exponential operator $\exp\{T\}$ to the ground configuration $\left|\phi_0\right>$.
In practice, one truncates of course the operator $T$.
For fermions, it is widely known that truncated coupled-cluster expansions
are size consistent, which is another advantage the coupled-cluster approach
possesses in comparison to truncated CI expansions which are not size consistent \cite{Szabo_book}.

Our aim in this work is to derive a coupled-cluster theory for bosons with
emphasis on systems of interacting indistinguishable bosons in traps with up to many particles. 
We investigate aspects like size consistency and what to use as the 
initial ground configuration $\left|\phi_0\right>$.
We would like to mention that coupled-cluster approaches for molecular vibrations \cite{Vib_CC},
``bosonic nuclei'' \cite{CC_bos_nuc}, the spin-boson model \cite{CC_spin_boson},
and within bosonization of many-electron systems \cite{Mukamel}
have been studied in the literature, 
but are very different from the present work.

The structure of the paper is as follows.
In section II, we briefly discuss the standard configuration-interaction approach.
In section III, the coupled-cluster theory for bosons is developed,
where the issue of size consistency is extensively analyzed.
Working equations for a truncation of the coupled-cluster to
single and double excitations (CCSD) are derived in section IV,
and an illustrative numerical example is provided in section V.
Finally, summary and conclusions are drawn in section VI.

\section{The usual approach: Configuration interaction}

Consider a system of interacting $N$ identical particles,
for simplicity either spinless or all of the same spin projection.
We introduce $M$ one-particle functions
$\varphi_i(\vec r), i=1,2,\ldots,M$, 
which are called orbitals.
The $N$ particles can be distributed over these orbitals and
each allowed distribution defines a configuration
$\Phi_{i_1,i_2,\ldots,i_N}$.
If the particles are fermions, the configuration is a determinant
\beq\label{eq1}
\Phi_{i_1,i_2,\ldots,i_N} = \hat {\cal A} \varphi_{i_1} \varphi_{i_2} \ldots \varphi_{i_N}
\eeq
and if they are bosons, it is a permanent
\beq\label{eq2}
\Phi_{i_1,i_2,\ldots,i_N} = \hat {\cal S} \varphi_{i_1} \varphi_{i_2} \ldots \varphi_{i_N}.
\eeq
$\hat {\cal A}$ and $\hat {\cal S}$
denote the antisymmetrizing and symmetrizing operators, respectively.
In the absence of interaction,
each configuration is an eigenfunction of the Hamiltonian $H$
of the system.
In the presence of interaction between the particles,
the exact eigenfunction $\Psi$ in the space defined by the $M$ orbitals
is given by a superposition of all the allowed configurations
\beq\label{eq3}
\Psi = \sum_{i_1,i_2,\ldots,i_N} D_{i_1,i_2,\ldots,i_N} \Phi_{i_1,i_2,\ldots,i_N},
\eeq
where the $D$'s are complex numbers.
The $D$'s are usually determined variationally by diagonalizing
the Hamiltonian matrix 
$\left\{ \left< \Phi_{i_1,i_2,\ldots,i_N} \left| H \right| \Phi_{j_1,j_2,\ldots,j_N} \right> \right\}$.
Clearly, to obtain the correct exact eigenfunction,
the orbital basis should be complete, i.e. $M \to \infty$, but in practical calculations
$M$ is kept finite.

How many distinct configurations participate in the configuration
interaction (CI) expansion (\ref{eq3})?
Two fermions cannot reside in a single orbital and, therefore,
the number of configurations is simply given by
\beq\label{eq4}
 {\cal F}^M_N = 
\left( M \atop N \right). 
\eeq
In the case of bosons there is no restriction on how many particles
can reside in an orbital.
We find that the number of bosonic configuration reads
\beq\label{eq5}
 {\cal B}^M_N = 
\left( M+N-1\atop N \right).
\eeq
These numbers grow rapidly with the size $M$ of the orbital basis
and much more rapidly for bosons than for fermions.
Consider, for example, $165$ particles.
For fermions $M=165$ is needed in order to have a single configuration.
Adding just $5$ more orbitals, i.e. $M=170$,
increases the number of configurations to over a billion ($10^9$).
For bosons, $M=1$ is needed to have a single configuration and
employing $M=170$ leads to an astronomically
large number of configurations.

For $165$ fermions to have only $5$ additional (so called virtual)
orbitals at their disposal is usually insufficient for
the calculation of their correlation energy.
For an accurate calculation more virtual orbitals are required making the
CI approach impractical.
Fortunately, the number of orbitals needed for accurate calculations
for bosons is much less than for fermions.
Because many or even all bosons may reside in a single orbital,
the structure of the orbitals used play a major role in the
calculation and the appropriate choice of the orbitals is essential.
The orbitals are preferentially determined self-consistently as
done, for instance, by the use of the GP equation \cite{Gross_Pitaevskii},
see also \cite{review_Legget,Stringari_book}.
Nevertheless, to achieve meaningful results $M$ is not small
and the number of configurations is often beyond reach.
To return to our example of $N=165$,
the number of configurations exceeds a billion with just $M=6$,
i.e., with just $5$ additional (virtual) orbitals.
Note that the numbers of bosonic and fermionic configurations
are identical for the same number of virtual orbitals
($M-N$ orbitals for fermions and $M-1$ for bosons)
as can be seen from Eqs.~(\ref{eq4}) and (\ref{eq5}).

In the following we concentrate on bosons and make use
of the destruction and creation operators $b_i$ and $b_i^\dag$,
$i=1,2,\ldots,M$,
corresponding to the orbitals $\varphi_i$ introduced above.
These operators fulfill the usual commutator relations
\beq\label{eq6}
 \left[b_i,b_j^\dag\right] = \delta_{i,j} \qquad \qquad 
  \left[b_i,b_j\right] = \left[b_i^\dag,b_j^\dag\right]=0
\eeq
for bosons.
Utilizing these operators, we define the ground configuration
\beq\label{eq7}
 \left|\phi_0\right> = \frac{1}{\sqrt{N!}} (b_1^\dag)^N \left|0\right>, \qquad
 \left<\phi_0\left|\right.\phi_0\right> = 1
\eeq
which is the ground state of the system in the absence 
of interaction between the particles.
$\left|0\right>$ denotes the vacuum.
All other configurations $\left|\Phi_{i_1,i_2,\ldots,i_N}\right>$
are obtained directly by applying excitation operators to $\left|0\right>$.
Singly excited configurations read $b_i^\dag b_1 \left|\phi_0\right>$,
doubly excited ones are given by $b_i^\dag b_j^\dag (b_1)^2 \left|\phi_0\right>$,
and so on.
In analogy to Eq.~(\ref{eq3}) the exact state $\left|\Psi\right>$ can be
expanded in these orthogonal configurations
\beqn\label{eq8}
\left|\Psi\right> = & &\left|\phi_0\right> +
 \sum_{i_1=2}^M d_{i_1} b_{i_1}^\dag b_1 \left|\phi_0\right> +
 \sum_{i_1,i_2=2}^M d_{i_1i_2} b_{i_1}^\dag b_{i_2}^\dag (b_1)^2 \left|\phi_0\right> \nonumber \\ 
 & & + \ldots + 
\sum_{i_1,i_2,\ldots,i_N=2}^M d_{i_1i_2\ldots i_N} 
 b_{i_1}^\dag b_{i_2}^\dag \ldots b_{i_N}^\dag (b_1)^N \left|\phi_0\right>. \
\eeqn
For later use we choose explicitly intermediate normalization of
the exact state, i.e., $\left<\phi_0\left|\right.\Psi\right> = 1$,
do not impose normalization on the orthogonal configurations except
on $\left|\phi_0\right>$, and allow for redundancy in that the
same configuration may appear several times in the expansion.
We note that the expansion coefficients are independent of the order of the
indices: $d_{i_1\ldots i\ldots j\ldots i_N} = d_{i_1\ldots j\ldots i\ldots i_N}$.
Obviously, there is a one to one correspondence between the coefficients in 
Eq.~(\ref{eq8}) and those in the expansion in distinct normalized configurations.

With the aid of the expansion (\ref{eq8}) it is relatively straightforward to
express the exact energy $E_0$ in terms of the expansion coefficients.
Starting from the Schr\"odinger equation $H\left|\Psi\right>=E_0\left|\Psi\right>$
one immediately arrives at
\beq\label{eq9}
 E_0 =  \left<\phi_0\left|H\right|\Psi\right>.
\eeq
As usual the system's Hamiltonian consists of an one-particle
operator $\hat h(\vec r)$ and a two-particle interaction $\hat V(\vec r - \vec r')$.
Expressed in destruction and creation operators $H$ takes on the common appearance \cite{Fetter_book}
\beq\label{eq10}
 H = \sum h_{ij} b_i^\dag b_j + \frac{1}{2} \sum V_{ijkl}  b_i^\dag  b_j^\dag  b_k  b_l
\eeq
where
\beqn\label{eq11}
 h_{ij} = & & \int \varphi_i^\ast \hat h \varphi_j d\vec r \, , \nonumber \\
 V_{ijkl} = & & \int\!\!\int \varphi_i^\ast(\vec r) \varphi_j^\ast(\vec r')
  \hat V(\vec r - \vec r') \varphi_k(\vec r) \varphi_l(\vec r')d\vec rd\vec r'  . \
\eeqn
Inserting Eqs.~(\ref{eq8}) and (\ref{eq10}) into (\ref{eq9}) leads to
\beq\label{eq12}
\left[E_0 - \left<\phi_0\left|H\right|\phi_0\right> \right]/N =
 \sum_{l=2}^M \left[h_{1l} + (N-1) V_{111l}\right] d_l +
 (N-1)\sum_{k,l=2}^M V_{11kl} d_{kl}.
\eeq
The energy correction per particle due to the dressing
$\phi_0 \rightarrow \Psi$ can be expressed by the coefficients 
$\{d_l\}$ and $\{d_{lk}\}$.
The orbitals $\{\varphi_l\}$ can be conveniently chosen to simplify 
Eq.~(\ref{eq12}) further by eliminating the $\{d_l\}$;
see next chapter for details.

We should keep in mind that in spite of the compactness
of expression (\ref{eq12}) this equation cannot be used to determine
the unknown coefficients $\{d_l,d_{kl}\}$ since $\left<\phi_0\left|H\right|\Psi\right>$
in Eq.~(\ref{eq9}) is not subject to a variational principle.
These coefficients are determined by diagonalizing the Hamiltonian matrix
which is -- as discussed above -- of immense dimensionality and in general
not amenable to practical calculations.
One, therefore, resorts to approximations like keeping $M$ very small
and/or truncating the CI expansion (\ref{eq8}).
A particularly appealing approach is to truncate the expansion by taking
into account only a few classes of configurations.
In analogy to electron structure calculations we may consider $\left|\phi_0\right>$
and all singly excited configurations (CIS),
add to these all doubly excited configurations (CISD), and so on.

\section{Coupled-cluster theory for bosons}

\subsection{General aspects}

The CI approach discussed in the preceding section is formally straightforward but impractical.
Truncating the CI expansion cannot be expected to solve the
problem satisfactorily in many cases.
In particular, when the interaction between the bosons is substantial
and/or many bosons are present, numerous highly excited configurations
may contribute rendering systematic truncations impossible.
We are, therefore, searching for more efficient approaches
which are amenable to systematic approximations.

In the coupled-cluster approach the exact wavefunction
is obtained by applying an exponential operator to the ground configuration (\ref{eq7}):
\beq\label{eq13}
\left|\Psi\right> = e^T \left|\phi_0\right>.
\eeq
The operator $T$ is a superposition of excitation operators
\beq\label{eq14}
 T = \sum_{n=1}^N T_n,
\eeq
where for bosons we may write
\beqn\label{eq15}
 T_n = & & t_n (b_1)^n \nonumber \\
 t_n = & & \sum^M_{i_1,\ldots,i_n=2} c_{i_1i_2\ldots i_n} 
 b_{i_1}^\dag b_{i_2}^\dag \ldots b_{i_n}^\dag .
\eeqn
For simplicity we have again introduced redundancies to avoid
unpleasant restrictions on the summation indices.
The yet unknown coefficients $c_{i_1i_2\ldots i_N}$ do not
depend on the ordering of the subscripts, i.e., $c_{i_1i_2}=c_{i_2i_1}$ etc.
It is convenient to note that $[T_n,T_m]$=0 
and hence $\exp(T)=\exp(T_N)\ldots\exp(T_1)$.
Because of the simple structure of $\left|\phi_0\right>$, see Eq.~(\ref{eq7}),
the appearance of the coupled-cluster operator $T$ for bosons is much 
simpler than that for fermions, see, e.g., \cite{CC_r3,rev1}.

Using Eq.~(\ref{eq13}) and the Schr\"odinger equation it is easily seen that the exact energy reads
\beq\label{eq16}
 E_0 = \left<\phi_0\left|e^{-T}He^{T}\right|\phi_0\right>.
\eeq
The wavefunction (\ref{eq13}) is subject to intermediate normalization
$\left<\phi_0\left|\right.\Psi\right> = 1$ as can be deduced directly from
$\left<\phi_0\right| \exp(\pm T) = \left<\phi_0\right|$.
In this respect the situation is similar to that discussed in the preceding chapter,
see Eq.~(\ref{eq9}).
On the other hand, Eq.~(\ref{eq16}) is much more powerful because
$\exp(-T)H\exp(T)$ can be evaluated using the useful expansion
\beq\label{eq17}
 \dot A \equiv e^{-T}Ae^{T} = A + 
\frac{1}{1!}\left[A,T\right]+\frac{1}{2!}\left[\left[A,T\right],T\right]+\ldots \, ,
\eeq
which can be applied to any operator $A$.

As discussed in the preceding chapter, an expression like (\ref{eq16})
is not subject to a variational principle and cannot be used to
determine the unknown coefficients $c_{i_1i_2\ldots i_n}$.
To proceed we notice that $e^{-T}He^{T}\left|\phi_0\right>=E_0\left|\phi_0\right>$
and hence projecting on any excited configuration provides an equation for the coefficients.
The singly excited configurations lead to the $(M-1)$-equations
\beq\label{eq18}
  \left<\phi_0\left|b_1^\dag b_i e^{-T}He^{T}\right|\phi_0\right> = 0  \qquad \qquad
  i=2,3,\ldots,M
\eeq
and the doubly excited ones to the $M(M-1)/2$ distinct equations 
\beq\label{eq19}
  \left<\phi_0\left|(b_1^\dag)^2 b_i b_j e^{-T}He^{T}\right|\phi_0\right> = 0  \qquad \qquad
  i\ge j=2,3,\ldots,M
\eeq
and so on.
The number of independent equations corresponds exactly
to the number of distinct coefficients, $M-1$ coefficients $c_i$,
$M(M-1)/2$ coefficients $c_{ij}$, etc.
The equations above are nonlinear in the unknown coefficients and,
furthermore, are coupled to each other.
The set (\ref{eq18}) contains the $c_i$ as well as the $c_{ij}$,
while the set (\ref{eq19}) also the $c_{ijl}$.
Since the highest possible excitation is $N$-fold,
the final set of equations does not contain new unknown coefficients.

The total number of distinct coefficients in Eq.~(\ref{eq15}) is, of course,
identical to the total number of distinct bosonic configurations
given in Eq.~(\ref{eq5}).
We have argued above that this number is enormous.
Moreover, the equations like (\ref{eq18}) used to determine
the coefficients are nonlinear.
So where is the gain with respect to the standard CI method discussed in the preceding chapter?
The gain is in the favorable properties of the coupled-cluster ansatz (\ref{eq13})
when truncating the sum of excitation operators in Eq.~(\ref{eq14}).
Let us for demonstration include only single and double excitation operators in $T$,
i.e., $T=T_1+T_2$.
Then, the only coefficients available are the $c_i$ and $c_{ij}$ which
can be determined from Eqs.~(\ref{eq18}) and (\ref{eq19}).
By inspecting that 
\beq\label{eq20}
 \left|\Psi\right> = \left|\phi_0\right> + (T_1+T_2)\left|\phi_0\right> +
 \frac{1}{2!} \left(T_1^2 + 2T_1T_2 + T_2^2\right)\left|\phi_0\right> +\ldots
\eeq
one readily notices that this expansion of the wavefunction contains
{\it all possible} distinct configurations of the system.
The Eqs.~(\ref{eq18}) and (\ref{eq19}) determine the $c_i$ and $c_{ij}$
coefficients such that the expansion (\ref{eq20}) is optimal in providing
$e^{-T}He^{T}\left|\phi_0\right>$. In contrast to this CCSD approach as we would 
call it in analogy, the CISD expansion, on the other hand,
knows only singly and doubly excited configurations,
i.e., is rather related to truncating (\ref{eq20}) as
$\left|\Psi\right> = \left|\phi_0\right> + \left(T_1 + \frac{1}{2!}T_1^2 + T_2\right)\left|\phi_0\right>$.
For additional advantages of the coupled-cluster ansatz see the following
two sections.

\subsection{Impact of the single excitation operator $T_1$}

The influence of $T_1$ is particularly transparent.
For this purpose we consider $\exp(T_1)\left|\phi_0\right>$ and
remind that the ground configuration is particularly simple for bosons,
see Eq.~(\ref{eq7}).
Using the series (\ref{eq17}), it is easily seen that
\beq\label{eq21}
 e^{T_1} b_1^\dag e^{-T_1} = b_1^\dag + \sum_{l=2}^M c_l b_l^\dag
\eeq
which defines a new creation operator
\beq\label{eq22}
 \tilde b_1^\dag = \frac{1}{\sqrt{1+|c|^2}} \left(b_1^\dag + \sum_{l=2}^M c_l b_l^\dag\right),
\eeq
where $|c|^2=\sum_{l=2}^M |c_l|^2$,
which fulfills the boson commutator relation
$[\tilde b_1,\tilde b_1^\dag]=1$.
Consequently, the action of $\exp(T_1)$ on the ground permanent $\phi_0$
is to define a new permanent
\beq\label{eq23}
 \left|\tilde \phi_0\right> \equiv e^{T_1} \left|\phi_0\right> \propto 
 \left(\tilde b_1^\dag\right)^N \left|0\right>
\eeq
which is, however, not normalized to $1$,
but rather to $\left<\tilde \phi_0\left|\right.\tilde\phi_0\right> = \left[1+|c|^2\right]^N$.

To proceed, one can consider the quantities appearing in Eq.~(\ref{eq22}) as the
first column of an unitary matrix $\mathbf U$ which defines a new set of $M$
creation operators $\left(\tilde b_1^\dag,\tilde b_2^\dag,\ldots,\tilde b_M^\dag\right)=
\left(b_1^\dag,b_2^\dag,\ldots,b_M^\dag\right) \mathbf U$.
This transformation defines a new set of corresponding orthonormal orbitals
$\tilde\varphi_1,\tilde\varphi_2,\ldots,\tilde\varphi_M$.
In turn, the new set of creation and destruction operators or,
equivalently, of orbitals, can be formally utilized to eliminate $T_1$ from $T$.
The remaining operators of $T$, the $T_n$ with $n\ge2$,
are now defined with the operators $\tilde b_1$ and $\tilde b_i^\dag$,
e.g., $T_2=\sum_{i_1,i_2=2}^M \tilde c_{i_1i_2}
\tilde b_{i_1}^\dag \tilde b_{i_2}^\dag (\tilde b_1)^2$.

Clearly, the impact of $T_1$ is to introduce a new orbital
$\tilde \varphi_1$ optimal for the coupled-cluster expansion.
In particular, if we put all $T_n=0, n\ge2$,
this new orbital can be constructed explicitly.
As discussed in chapter IV,
this orbital then minimizes the energy functional 
$\left<\phi_0\left|H\right|\phi_0\right>$.

\subsection{Size consistency}

Let us consider a super system consisting of $R$ noninteracting
replica of our original $N$-particle system.
Clearly, the exact energy of this super system is $E_0(R)=R\cdot E_0$,
where $E_0$ is the energy of the $N$-particle system.
This result will, of course, be reproduced if either the full configuration
interaction expansion (\ref{eq8}) or the coupled-cluster expansion (\ref{eq13}-\ref{eq15}) is used.
In general, the full expansion cannot be utilized and one has to resort to approximations.
We, therefore, have to pose the question whether truncated CI and coupled-cluster expansions
for bosonic systems lead to energies which scale correctly with the number of
replica $R$, i.e., whether these truncated expansions are {\it size consistent}.

Size consistency plays as important role in electronic structure calculations \cite{Szabo_book}.
Imagine, for instance, a molecule which is being broken up into fragments
or a cluster consisting of weakly interacting atoms.
The computational methods used must be size consistent in order
to describe correctly the break up of the molecule into fragments
or the cluster.
Indeed, it is well known that truncated CI expansions are generally
{\it not} size consistent whereas truncated CC expansions are size
consistent for electrons.
In the following we would like to address the issue of size consistency for bosons.
The concept of size consistency is also relevant for bosons.
Bosonic systems, e.g., in an external double-well trap can be fragmented \cite{ALN_PRA,Fer},
and the computational method used must be able to describe fragmentation correctly.
Another, even more extreme example is the superfluid
to Mott-insulator transition of Bose-Einstein condensates in a lattice trap \cite{Jaksch1_PRL,IB1_nature}.
In the superfluid phase all bosons communicate with each other 
and in the insulator phase each potential well of the lattice contains a
single boson which hardly interacts with the other bosons.

The ground configuration $\phi_0$ of the super system is a symmetrized
product of the $R$ ground configurations of the individual replica. We write
\beq\label{eq24}
 \left|\phi_0\right> = \frac{1}{(N!)^{R/2}} \prod_{k=1}^R \left(b_{1_k}^\dag\right)^N \left|0\right>
\eeq
where $b_{1_k}^\dag$ is the creation operator for bosons in
the occupied orbital of the $k$-th replica.
The Hamiltonian of the super system is, of course, just the sum of the
individual Hamiltonians
\beq\label{eq25}
 H = H_1 + H_2 + \ldots + H_R.
\eeq

We first show that the truncated CI expansion is not size consistent.
For this purpose we proceed in analogy to the considerations done
for fermions (electrons) \cite{Szabo_book} and assume each of the $R$ replica to consist of two orbitals,
or equivalently two destruction operators $b_{1_k}$ and $b_{2_k}$,
of different spatial symmetry.
It is sufficient to demonstrate that CID is not size consistent.
This implies that the expansion of the total wavefunction $\left|\Psi\right>$
consists of the superposition of the ground configuration $\left|\phi_0\right>$
given above in Eq.~(\ref{eq24}) and of the doubly excited configurations
$\left(b_{2_k}^\dag\right)^2\left(b_{1_k}\right)^2 \left|\phi_0\right>, k=1,2,\ldots,R$.
In this space the Hamiltonian matrix $\bH$ representation of
$H$ is an ``arrow'' matrix of dimension $R+1$, the elements of which read
\beqn\label{eq26}
 {\mathcal H}_{00} = & & \left<\phi_0\left|H\right|\phi_0\right>  \nonumber \\
 {\mathcal H}_{0k} = & & C\left<\phi_0\left|H_k\left|
\left(b_{2_k}^\dag\right)^2\left(b_{1_k}\right)^2\right.\right|\phi_0\right>  \nonumber \\
 {\mathcal H}_{kk'} = & & |C|^2\left<\phi_0\left|\left(b_{1_k}^\dag\right)^2\left(b_{2_k}\right)^2
\right|H_k\left|\left(b_{2_k}^\dag\right)^2\left(b_{1_k}\right)^2\right|\phi_0\right> \delta_{kk'}, \
\eeqn
where $C$ is the normalization constant of a double excited configuration.
Note that all matrix elements
${\mathcal H}_{0k}, k=1,2,\ldots,R$ are identical to each other and so are all the ${\mathcal H}_{kk}$.
We put for convenience ${\mathcal H}_{0k}=V$ and 
${\mathcal H}_{kk}= \left<\phi_0\left|H\right|\phi_0\right>+\Delta$.
The diagonalization of $\bH$ can be performed analytically by searching for the roots of
$E-\left<\phi_0\left|H\right|\phi_0\right> = \sum_k {\left|{\mathcal H}_{0k}\right|}^2 / 
\left[E-{\mathcal H}_{kk}\right]$.
This immediately leads to
\beq\label{eq27}
 E_0(R) = \left<\phi_0\left|H\right|\phi_0\right>+\frac{\Delta}{2} - R^{1/2} 
\left[V^2 + \frac{\Delta^2}{4R}\right]^{1/2}
\eeq
which implies that the truncated CI expansion is not size consistent.
Using Eq.~(\ref{eq24}) and (\ref{eq25}) one sees that the expectation value
$\left<\phi_0\left|H\right|\phi_0\right>$ is size consistent and the correction
term $E_0 - \left<\phi_0\left|H\right|\phi_0\right>$ scales as $R^{1/2}$
for large $R$ instead of being proportional to $R$.

In contrast to the truncated CI expansion, the truncated coupled-cluster expansion is size consistent.
The operator $T$ is a sum of $T^{(k)}$ for the $k=1,2,\ldots,R$ replica.
Each of the $T^{(k)}$ has the appearance as in Eqs.~(\ref{eq14}) and (\ref{eq15})
for the individual replica. One has just to index the destruction and annihilation operators
appearing there by a further subscript $k$ for the $k$-th replica.
The values of the coefficients $c$ in Eq.~(\ref{eq15}) are, of course,
the same for all replica.
Clearly, the various $T^{(k)}$ commute with each other and, consequently, $\exp\{T\}$
can be factorized as $\prod_{k=1}^R \exp\{T^{(k)}\}$ leading to
\beq\label{eq28}
 \left|\Psi\right> = \left[e^{T^{(1)}} \left(b_{1_1}^\dag\right)^N \right]
                     \left[e^{T^{(2)}} \left(b_{1_2}^\dag\right)^N \right] \ldots
                     \left[e^{T^{(R)}} \left(b_{1_R}^\dag\right)^N \right] \left|0\right>
\eeq
which is size consistent for any truncation of the $T^{(k)}$.

In spite of the favorable structure (\ref{eq28}) a major problem arises.
If we {\it a priori} know that our system consists of $R$ noninteracting replica,
we may, of course, use the $\phi_0$ in Eq.~(\ref{eq24}) and obtain a size consistent result.
However, the intension is to apply the coupled-cluster method not knowing {\it a priori}
how our system behaves, i.e., whether it is superfluid or breaks up
into weakly interacting subsystems.
Lacking this knowledge, we cannot use the ansatz (\ref{eq24}) for $\phi_0$.
Resorting to
\beq\label{eq29}
  \left|\phi_0\right> = \frac{1}{\sqrt{(NR)!}}\left(b_1^\dag\right)^{NR} \left|0\right>
\eeq
which does not distinguish between the $R$ replica as is the case in Eq.~(\ref{eq24}),
we may again pose the question: is a truncated coupled-cluster ansatz size consistent?

To proceed, we first have to identify the $b_1^\dag$ operator appearing in Eq.~(\ref{eq29})
in terms of the operators $b_{1_k}^\dag$ of the individual replica.
Since all replica are identical,
we can construct $R$ new operators $B^\dag_{11},B^\dag_{12},\ldots,B^\dag_{1R}$
of the super system by linearly combining the $b_{1_k}^\dag$.
Without loss of generality we can always chose
\beq\label{eq30}
 b_1^\dag \equiv B_{11}^\dag = R^{-1/2} \left(b_{1_1}^\dag+b_{1_2}^\dag+\ldots+b_{1_R}^\dag\right),
\eeq
i.e., as a trivial superposition of the creation operators
corresponding to the occupied orbitals of the individual replica.
All the $B_{1i}^\dag$ will posses different permutational symmetries which 
simplifies the evaluation considerably. For instance, for $R=2$ we have
$B_{11}^\dag=2^{-1/2} \left(b_{1_1}^\dag+b_{1_2}^\dag\right)$ and
$B_{12}^\dag=2^{-1/2} \left(b_{1_1}^\dag-b_{1_2}^\dag\right)$.
We note that for each set of virtual orbitals
an analogous procedure can be applied to introduce the remaining orbitals
of the super system: $b_{2_1}^\dag,b_{2_2}^\dag,\ldots,b_{2_R}^\dag$
are linearly combined to give $B_{21}^\dag,B_{22}^\dag,\ldots,B_{2R}^\dag$ and so on.
This results in $R\cdot M$ creation operators of the super system emerging from the
$M$ operators of each of the replica.
Since only one orbital is occupied in the ground configuration of the super systems,
all the other ones are virtual orbitals, i.e., also
the $B_{12}^\dag,B_{13}^\dag,\ldots,B_{1R}^\dag$ refer now to virtual orbitals.

The Hamiltonian (\ref{eq25}) and the coupled-cluster operator $T$ are now expressed in the
$B_{ik}^\dag$ of the super system. Let us consider as an example the one-body
part of $H$ in the occupied space of the individual replica:
$$
 \sum_k^R h_{11} b_{1_k}^\dag b_{1_k} = \sum_k^R h_{11} B_{1k}^\dag B_{1k}.
$$
Note that in the two-body part of the $H$ operator 
products like $B_{1k}^\dag B_{1k'}^\dag B_{1k} B_{1k'}, k\ne k'$, appear.
Let us begin the analysis by inspecting the mean-field energy
$\left<\phi_0\left|H\right|\phi_0\right>$.
Here, only the terms of the Hamiltonian
containing $B_{11}^\dag B_{11}$ and 
$B_{11}^\dag B_{11}^\dag B_{11} B_{11}$ contribute.
These terms take on the explicit appearance
$$
  h_{11}B_{11}^\dag B_{11} + \frac{V_{1111}}{2R}B_{11}^\dag B_{11}^\dag B_{11} B_{11}  
$$
where $h_{11}$ and $V_{1111}$ are the quantities defined in Eq.~(\ref{eq11}) for an
individual replica.
One immediately finds
\beqn\label{eq31}
 \left<\phi_0\left|H\right|\phi_0\right> = & & NRh_{11} + \frac{NR(NR-1)}{2R} V_{1111} \nonumber \\
 = & & R \left\{ Nh_{11} + \frac{N\left(N-1/R\right)}{2} V_{1111} \right\} 
\eeqn
implying that even the mean-field energy is not size consistent; a surprising result.
The mean-field energy of an individual replica is $Nh_{11} + \frac{N(N-1)}{2} V_{1111}$.
Consequently, size consistency is achieved only if each individual replica
contains many bosons, i.e., for $N\gg 1$.

To better understand the implications of the above finding,
let us briefly consider the coupled-cluster operator $T$
of the super system. Since now there is only a single occupied orbital
(related to $B_{11}$), {\it all} the other operators 
$B_{ik}^\dag$ relate to virtual orbitals of the super system.
Consequently, $T$ can be broken up into a part $T'$ which contains excitations solely within the 
original occupied orbitals of the different replica 
and the remaining part $T''$
where excitations to the originally virtual orbitals of these replica are included.
As an example we consider the double excitation operator $T_2$ (see Eqs.~(\ref{eq14}) and (\ref{eq15})):
\beq\label{eq32}
 T_2 = T'_2 + T''_2 = \sum_{k=2}^R c_k \left(B_{1k}^\dag\right)^2 \left(B_{11}\right)^2
 + \sum_{k=1}^R \sum_{i,j}^M \!' c_{ijk} B_{ik}^\dag B_{jk}^\dag \left(B_{11}\right)^2.
\eeq
In $T''_2$ the terms with $B_{11}$ and those of $T'$ 
are not included as indicated by the primed summation symbol $\sum'$.
In the example of two replica, we have the $T_2'$ excitation operator 
$\left(b_{1_1}^\dag-b_{1_2}^\dag\right)^2\left(b_{1_1}+b_{1_2}\right)^2$
which is actually an excitation within the occupied manifold of the replica.
Obviously $T'$ and $T''$ commute.

Interestingly, the full impact of $\exp\{T'\}$ is needed in order to restore the size consistency.
Indeed, a calculation shows that
\beq\label{eq32t}
\left<\phi_0\left|e^{-T'}He^{T'}\right|\phi_0\right> =
R \left\{N h_{11} + \frac{N(N-1)}{2}V_{1111}\right\}
\eeq
which is the expected correct mean-field result and is identical to the expectation value of $H$
obtained with the ansatz (\ref{eq24}) for $\phi_0$ 
where the knowledge of having $R$ replica has been used.

The result (\ref{eq32t}) follows only if the expansion of $\exp\{T'\}$ is fully considered and not truncated.
The impact of $\exp\{T'\}$ is to transform $\phi_0$ in Eq.~(\ref{eq29}) into the form of 
Eq.~(\ref{eq24}) which is appropriate for $R$ replica.
In other words, truncated coupled-cluster expansions are not size consistent once the ansatz (\ref{eq29})
is used for $\phi_0$. 
The good news is that the violation of the size consistency 
at least for the mean-field energy
leads to negligible errors for large
individual systems ($N\gg 1$), see Eq.~(\ref{eq31}).
In this respect bosons and fermions behave differently.
Due to the fact that each fermion resides in its own orbital, size consistency in truncated
coupled-cluster expansions follows straightforwardly, as was also
found above for bosons starting with the $\phi_0$ of Eq.~(\ref{eq24}).

\subsection{On the choice of the ground configuration $\phi_0$}

In contrast to fermions, the choice of the structure of the ground
configuration $\phi_0$ as the starting point is crucial for bosons if fragmentation
or, in particular, phase transitions like the superfluid to Mott-insulator
transition are to be studied.
In the absence of interaction between the bosons, the exact ground state has the
appearance $\propto \left( b_1^\dag\right)^N \left|0\right>$.
It is, therefore,
natural to start in the presence of interaction from an analogously structured ground
configuration $\phi_0$ as done in what follows Eq.~(\ref{eq7}).
In the presence of interparticle interaction we have the freedom to
choose the orbitals defining the destruction and annihilation operators.
At least as long as this interaction is weak, it is favorable to choose the occupied
orbital which minimize the energy expectation value $\left<\phi_0\left|H\right|\phi_0\right>$.
This readily leads to the equation
\beq\label{eq33}
 \left[ \hat h + (N-1) \hat J_{11} \right] \varphi_1(\vec r) = \mu_1 \varphi_1(\vec r)  
\eeq
which determines the occupied orbital $\varphi_1(\vec r)$.
The number $\mu_1$ can be called orbital energy or chemical potential.
The direct interaction operator $\hat J_{11}$ is a local operator and reads
\beq\label{eq34}
 \hat J_{11} = \int \varphi_1^\ast(\vec r') \hat V(\vec r - \vec r') \varphi_1(\vec r') d\vec r'.
\eeq
Eq.~(\ref{eq33}) defines an hermitian Fock-like operator
\beqn\label{eq35}
 \hat F \equiv & & \hat h + (N-1) \hat J_{11} \nonumber \\
 \hat F \varphi_i = & & \mu_i \varphi_i \
\eeqn
the eigenfunctions of which define a complete set of orthogonal orbitals
to be used in the coupled-cluster calculation.

For convenience (see chapter IV) one may introduce the more physical operator $\hat{\bar F}$
\beqn\label{eq36}
 \hat{\bar F} \equiv & & \hat h + \frac{N-1}{2} \left[\hat J_{11} + \hat K_{11}\right] \nonumber \\
 \hat{\bar F} \varphi_i = & & \mu_i \varphi_i \
\eeqn
which also contains the nonlocal exchange interaction operator $\hat K_{11}$:
\beq\label{eq37}
 \hat K_{11} \varphi_i = 
\int \varphi_1^\ast(\vec{r}') \hat V(\vec r - \vec{r}') \varphi_i(\vec{r}') \varphi_1(\vec r) d\vec{r}'.
\eeq
Because of the structure of $\phi_0$,
both $\hat F$ and $\hat{\bar F}$ produce the same
occupied orbital $\varphi_1$ and the same chemical potential.
All other orbitals and orbital energies are generally different.
To avoid confusion, we shall indicate in the following which set of orbitals
has been used.
Finally, we would like to point out that if one chooses 
$\hat V(\vec r - \vec r') \propto \delta(\vec r - \vec r')$, both
$\hat F \varphi_1 = \mu_1 \varphi_1$ and $\hat {\bar F} \varphi_1 = \mu_1 \varphi_1$
reduce to the well-known and widely used GP equation \cite{review_Legget,Stringari_book}.

As long as the system does not undergo a break up like
in the superfluid to Mott-insulator transition in an optical lattice potential,
$\phi_0$ of Eq.~(\ref{eq7}) and the orbital set of Eq.~(\ref{eq35}),
or, preferentially, of Eq.~(\ref{eq36}) can be used in the coupled-cluster calculations.
What to do when a break up is possible?
Here, we would like to stress that Eq.~(\ref{eq33}) has been obtained from the minimization of the
mean-field energy $\left<\phi_0\left|H\right|\phi_0\right>$ within the ansatz (\ref{eq7}) for $\phi_0$.
But, this ansatz does not necessarily lead to the lowest possible mean-field energy, i.e., 
it is not necessarily the best mean-field ansatz.
The best mean-field ansatz allows the bosons to reside in different orbitals \cite{LA_OAL_PLA}:
\beq\label{eq38}
\left|\phi_0\right>\propto\left(b_r^\dag\right)^{n_r}\!\!\ldots\left(b_2^\dag\right)^{n_2}
\left(b_1^\dag\right)^{n_1}\left|0\right>, \qquad \qquad
 n_1+n_2+\ldots n_r = N.
\eeq
The number $r$ of different orbitals as well as the occupation numbers $n_i,\, i=1,2,\ldots,r$
which tell us how many bosons reside in which orbital, are not {\it a priori}
fixed numbers but are determined variationally to minimize the mean-field energy.
The $r$ optimal orbitals involved are, of course, also determined variationally.
For brevity of presentation, we do not present the equations of
the best mean-field approach and refer to the literature \cite{LA_OAL_PLA}.

The best mean-field ansatz has been shown to be flexible
enough to predict and describe fragmentation and superfluid and a whole 
zoo of insulator phases \cite{ALN_PRA,Fer,Zoo}.
We, therefore, have reason to expect that (\ref{eq38}) provides 
a useful starting point for many coupled-cluster studies.
Other ways to determine the orbitals and their occupation numbers can
also be anticipated in connection with the coupled-cluster approach.

\section{Derivation of the working equations}

In this chapter the working equations of the coupled-cluster approach are derived and discussed.
We concentrate here on the ansatz $\frac{1}{\sqrt{N!}} (b_1^\dag)^N\left|0\right>$
for the ground configuration for which all the necessary ingredients have been introduced
and discussed in the preceding chapter.
Working equations can also be derived starting from ansatz (\ref{eq38}) for $\phi_0$.
This ground configuration contains several occupied orbitals and consequently the working equations
are more elaborate.

We begin by transforming the boson destruction and creation operators with $\exp\{T\}$. 
Using the expansion (\ref{eq17}) one readily finds that
($\dot A \equiv e^{-T}Ae^T$) 
\beqn\label{eq39}
& & \dot b_1 = b_1 \nonumber \\
& & \dot b_i^\dag = b_i^\dag, \qquad i=2,3,\ldots,M. \
\eeqn
The destruction operator corresponding to the orbital occupied
in $\phi_0$, and the creation operators of the virtual orbitals are invariant
to the coupled-cluster transformation.
In contrast, the respective dual operators change
\beqn\label{eq40}
& & \dot b_1^\dag = b_1^\dag - {\cal L}_1 \nonumber \\
& & \dot b_i = b_i + {\cal L}_i \
\eeqn
where
\beqn\label{eq41}
& & {\cal L}_1 = \sum_{n=1}^N n t_n \left(b_1\right)^{n-1} \nonumber \\
& & {\cal L}_i = \sum_{n=1}^N n t_n^{(i)} \left(b_1\right)^{n}. \
\eeqn
The operators $t_n$ can be found in Eq.~(\ref{eq15}) and the operators $t_n^{(i)}$
operate in the virtual space and read
\beqn\label{eq42}
& & t_n^{(i)} = \sum^M_{i_2,i_3,\ldots,i_n=2} c_{i,i_2,i_3,\ldots,i_n} 
 b_{i_2}^\dag b_{i_3}^\dag \ldots b_{i_n}^\dag  \nonumber \\
& & t_1^{(i)} = c_i. \
\eeqn
In the calculations below it is gratifying to note that the $\cal L$ operators commute
\beq\label{eq43}
\left[{\cal L}_i,{\cal L}_j\right]=\left[{\cal L}_1,{\cal L}_i\right]=0
\eeq
and that their action on $\left<\phi_0\right|$ from the right is simple:
\beqn\label{eq44}
& & \left<\phi_0\right|{\cal L}_1=0, \qquad  \left<\phi_0\right|(b_1)^m{\cal L}_1=0, \nonumber \\
& & \ \left<\phi_0\right|{\cal L}_i=c_i \left<\phi_0\right|b_1. \
\eeqn

To proceed, we break up the Hamiltonian (\ref{eq10}) into several terms
according to the number of operators related to the occupied orbital $\varphi_1$.
The transformed one-body part $\dot H_0$ of the Hamiltonian then consists of four terms
\beq\label{eq45}
\dot H_0 = h_{11} \dot b_1^\dag b_1 +
         \sum_{k=2}^M h_{1k} \dot b_1^\dag \dot b_k +
         \sum_{k=2}^M h_{k1} b_k^\dag b_1 +
         \sum_{k,l=2}^M h_{kl} b_k^\dag \dot b_l
\eeq
out of which the second is the most involved one.
The transformed two-body operator $\dot V$ contains many contributions
which can be casted into nine terms which, for ease of presentation,
are listed in the Appendix.

We now calculate the energy $E_0 = \left<\phi_0\left|\dot H\right|\phi_0\right>$, see Eq.~(\ref{eq16}).
The first term of $\dot H_0$ in Eq.~(\ref{eq45}) and that of $\dot V$ in the Appendix
contribute because of Eq.~(\ref{eq44}) only to the mean-field energy giving
\beq\label{eq46}
 \left<\phi_0\left|H\right|\phi_0\right> = N\left[h_{11} + \frac{N-1}{2}V_{1111}\right].
\eeq
The only terms contributing to the energy correction
$E_0 - \left<\phi_0\left|H\right|\phi_0\right>$ are the second term of $\dot H_0$ in
Eq.~(\ref{eq45}) and the second and forth terms of $\dot V$ in the Appendix.
The final result for the exact energy reads
\beq\label{eq47}
E_0 = \left<\phi_0\left|H\right|\phi_0\right> + 
N\left\{\sum_{k=2}^M \left[h_{1k} + (N-1)V_{111k}\right]c_k + 
\frac{N-1}{2} \sum_{k,l=2}^M V_{11kl} \left(2c_{kl}+c_kc_l\right)\right\}.
\eeq
Inspection of this expression makes clear from which of the above mentioned
contributing terms the various matrix elements originate.
The appearance of the result (\ref{eq47}) is similar to that derived by the configuration interaction 
approach, see Eq.~(\ref{eq12}).
The major difference is in the $c_kc_l$ term which is missing in the CI expression (\ref{eq12})
and arises due to the contribution of the single excitation operator $T_1$ to the
wavefunction, see Eq.~(\ref{eq20}).

Until now the orbitals used are arbitrary and have not been specified.
If we utilize the optimized orbitals arising from the Fock-like operators $\hat F$ and $\hat{\bar F}$
discussed in section III.D, we obtain in {\it both cases} the same results for the exact energy:
\beq\label{eq48}
 E_0 = \left<\phi_0\left|H\right|\phi_0\right> + 
\frac{N(N-1)}{2} \sum_{k,l=2}^M V_{11kl} \left(2c_{kl}+c_kc_l\right).
\eeq  
The other term in Eq.~(\ref{eq47}) has disappeared due to the fact that
$\left<\varphi_k\left|\hat F\right|\varphi_1\right>=0$, see Eqs.~(\ref{eq35}) and (\ref{eq36}).
In analogy to the notion of electron correlation energy \cite{Szabo_book} we might call
the correction $E_0 - \left<\phi_0\left|H\right|\phi_0\right>$, which is caused by the interparticle
interaction beyond the mean field, boson correlation energy.

To determine the coefficients $\{c_{kl}\}$ and $\{c_k\}$ we have to evaluate the series of coupled
equations (\ref{eq18}),(\ref{eq19}) and so on as discussed in section III.A.
The series consists of $N$ sets of such equations, one set for each type of excitation operator 
$T_n, n=1,2,\ldots,N$.
In practical calculations the expansion $T=\sum T_n$ is truncated.
For instance, if $T_1$ and $T_2$ are considered and the $T_n, n\ge 3$, are put to zero,
then only the sets of equations (\ref{eq18}) and (\ref{eq19}) must be considered in order
to determine the derived coefficients.
In the following we calculate this CCSD approach as we may call it.

Whereas the expression (\ref{eq48}) for the energy is invariant to the choice of
either $\hat F$ or $\hat{\bar F}$ to define the orbital basis used,
the equations determining the coefficients do depend on this choice.
We have computed these equations for an arbitrary set of orthonormal orbitals,
but present here only the results obtained with the orbitals of $\hat{\bar F}$.
Let us begin with $\left<\phi_0\left|b_1^\dag b_i \dot H\right|\phi_0\right>=0$.
From the nine terms of $\dot V$ shown in the Appendix it is easy to see that
the fifth, eight and ninth terms do not contribute as they all
exhibit two creation operators for virtual orbitals.
All other terms contribute.
Using Eqs.~(\ref{eq40}-\ref{eq44}) we obtain a set of $M-1$ coupled equations ($i=2,3,\ldots,M$)
\beqn\label{eq49}
& & \left(\mu_1 - \mu_i\right)c_i = \sum_{k=2}^M h_{1k} \left(2c_{ki} + c_kc_i\right) +
(N-1)\Bigg\{\frac{1}{2} \sum_{k=2}^M \left(V_{1i1k} + V_{1ik1}\right)c_k + \nonumber \\
& & \!\! \sum_{k,l=2}^M \left(V_{1ikl} - V_{11kl}c_i\right)\left(2c_{kl} + c_kc_l\right) 
+ (N-2) \sum_{k,l=2}^M V_{11kl} \left(3c_{kli} + c_kc_{li} +c_lc_{ki}\right)\Bigg\}. \ 
\eeqn
This set of equations is the result of the exact evaluation of 
$\left<\phi_0\left|b_1^\dag b_i \dot H\right|\phi_0\right>=0$ and thus contains
the coefficients $c_{kli}$ of $T_3$.
These coefficients have to be put equal to zero if CCSD is to be evaluated.
Consulting section III.B we see that $c_k\ne0$ implies the introduction of a new optimized orbital
and we may assume that in CCS and $M\to\infty$ this orbital is just the eigenfunction $\varphi_1$
of $\hat F$ (or, equivalently, $\hat{\bar F}$).

To complete the CCSD we have to solve also for the set of $M(M-1)/2$ distinct
coupled equations resulting from 
$\left<\phi_0\left|\left(b_1^\dag\right)^2 b_ib_j \dot H\right|\phi_0\right>=0$,
see text around Eq.~(\ref{eq19}).
Here, all terms of $\dot H$ contribute except of the third term of $\dot H_0$ in Eq.~(\ref{eq45}).
Using the relations (\ref{eq40}-\ref{eq44}) and the expressions of $\dot H$ given in Eq.~(\ref{eq45})
and in the Appendix, the derivation of the coupled equations is lengthy but straightforward.
In principle, one could derive diagrammatic rules to simplify the procedure in
analogy to the situation for fermions \cite{CC_r3,CC_r4},
but this is unnecessary for bosonic systems, at least as
long as $\phi_0$ in Eq.~(\ref{eq7}) is used.
The resulting set of coupled equations reads ($i,j=2,3,\ldots,M$):
\beqn\label{eq50}
& & 2\left(2\mu_1 - \mu_i - \mu_j\right)c_{ij} - V_{ij11} = 
\sum_{k=2}^M \left(V_{ijk1} + V_{jik1}\right)c_k - \left(V_{i111}c_j + V_{j111}c_i\right) + \nonumber \\
& & \sum_{k=2}^M  \left[\left(V_{1i1k} + V_{1ik1}\right)\alpha_{kj} + 
\left( i \leftrightarrow j \right)\right] +
 \sum_{k}^M V_{111k}\beta_{kij} + \sum_{k,l=2}^M V_{11kl}\,\gamma_{klij} - \\
& & \sum_{k,l=2}^M \left(V_{1ikl}c_j+V_{1jkl}c_i\right)c_kc_l
+ \sum_{k,l=2}^M V_{ijkl} \left(2c_{kl} + c_kc_l\right). \nonumber \
\eeqn
In contrast to Eq.~(\ref{eq49}) which contains $c_{kli}$ coefficients arising from $T_3$,
we have concentrated in Eq.~(\ref{eq50}) on CCSD and put all coupled-cluster operators
$T_n, n\ge3$, to zero.
The quantities $\alpha,\beta$ and $\gamma$ appearing in Eq.~(\ref{eq50}) are given by
\beqn\label{eq51}
 & & \alpha_{kj} = c_{kj}(N-3) - c_kc_j  \nonumber \\
 & & \beta_{kij} = 2\left[\left(c_{kj}c_i+c_{ki}c_j\right)(3N-5) +
                   2c_{ij}c_k(N-2) + c_ic_jc_k\right] \\
 & & \gamma_{klij} = 4c_{ki}c_{lj}(N-2)(N-3)-2(N-2)\bigg[2c_{ij}\left(2c_{kl}+c_kc_l\right)+
                     c_{ki}c_lc_j+c_{li}c_kc_j+ \nonumber \\ 
 & & c_{kj}c_lc_i+c_{lj}c_kc_i\bigg] -\left(2c_{kl}+c_kc_l\right)\left(2c_{ij}-c_ic_j\right). \nonumber \
\eeqn
It is worth noting that the equations (\ref{eq49}) arising from 
$\left<\phi_0\left|b_1^\dag b_i \dot H\right|\phi_0\right>=0$ are all homogeneous, 
whereas the equations (\ref{eq50}) originating from
$\left<\phi_0\left|\left(b_1^\dag\right)^2 b_ib_j \dot H\right|\phi_0\right>=0$
are inhomogeneous.
The inhomogeneity $V_{ij11}$ is due to the fifth term of $\dot V$ given in the Appendix,
i.e., from the only term which is invariant to the $\exp(T)$ transformation.

Before closing this chapter let us briefly discuss CCS.
Here, we have to put in Eq.~(\ref{eq49}) all the $c_{kli}=0$ as well as all the $c_{kl}=0$
and disregard the set of equations (\ref{eq50}).
The resulting equations are homogeneous in the $c_k$ coefficients and $c_k=0$ is a proper solution.
This implies that CCS leads to that mean-field energy which is the minimum of 
$\left<\phi_0\left|H\right|\phi_0\right>$, see section III.D. 
Would we have not used the orbitals of $\hat{\bar F}$ but rather some set of arbitrary orthonormal orbitals,
then Eq.~(\ref{eq49}) will become an inhomogeneous equation and the $c_k\ne0$.

\section{Illustrative example}

As an example we apply the CCSD approach to $N$ interacting bosons in an external trap
and restrict the orbital space to two orbitals $\varphi_1$ and $\varphi_2$ of different
spatial symmetry.
Consequently, the CCSD (or equivalently the CCD) wavefunction reads
$\left|\Psi\right> = \exp\left[c_{22}\!\left(b_2^\dag\right)^2\!\left(b_1\right)^2\right]\left|\phi_0\right>$.
In other words, the wavefunction depends only on a single unknown parameter $c_{22}$.
It is easily seen that
\beq\label{eq52}
 \left|\Psi\right> = \sum_{m=0}^{N/2} \frac{c_{22}^m}{m!} \left[\frac{N!(2m)!}{(N-2m)!}\right]^{1/2}
 \left|N-2m,2m\right>,
\eeq
where for simplicity we assume $N$ to be an even number and $\left|m,m'\right>$ is the normalized configuration
with $m$ bosons in $\varphi_1$ and $m'$ bosons in $\varphi_2$.

We remind that in coupled-cluster theory intermediate normalization of the
wavefunction is used, $\left<\phi_0\left|\right.\Psi\right>=1$, and define the norm of the wavefunction
\beq\label{eq53}
 {\mathcal N} = \left<\Psi\left|\right.\Psi\right>.
\eeq
Using Eq.~(\ref{eq52}) it is readily shown that this norm obeys a local ``decay'' law as a function
of the parameter $c_{22}$:
\beq\label{eq54}
\frac{d{\mathcal N}}{dc_{22}} = \frac{\left<n_2\right>}{c_{22}} \cdot {\mathcal N}
\eeq
where $\left<n_2\right>$ is the expectation value of the occupation number of bosons in orbital $\varphi_2$.
Because of the different spatial symmetry of $\varphi_1$ and $\varphi_2$, these orbitals are 
 the eigenfunctions of the reduced one-particle density matrix (natural orbitals)
and the respective eigenvalues are $\left<n_1\right>$ and $\left<n_2\right>$
with $\left<n_1\right>+\left<n_2\right>=N$.  
Clearly,
\beq\label{eq55}
 \left<n_2\right> =  \left<\Psi\left|b_2^\dag b_2\right|\Psi\right> / \left<\Psi\left|\right.\Psi\right>.
\eeq
which can be evaluated by using Eq.~(\ref{eq52}).
Analogously, the variance of the occupation number of bosons in orbital $\varphi_2$
can be obtained from the second derivative of the norm
\beq\label{eq56}
\frac{d^2{\mathcal N}}{dc_{22}^2} = \frac{\left<n_2^2\right>-\left<n_2\right>}{c_{22}^2} \cdot {\mathcal N}
\eeq
where 
$\left<n_2^2\right> =  \left<\Psi\left|{(b_2^\dag b_2)}^2\right|\Psi\right> / 
\left<\Psi\left|\right.\Psi\right>$.
In the absence of interaction between the bosons, $c_{22}=0$ and 
$\left<n_2\right>=0$, ${\mathcal N}=1$.
As the interaction strength grows, the value of the coupled-cluster coefficient 
$\left|c_{22}\right|$ grows as well and with it the mean number of bosons in the orbital $\varphi_2$. 
The quantity $\left|\left<n_2\right>/c_{22}\right|$ increases and determines the rate of change of the
norm according to Eq.~(\ref{eq54}).

To be specific, we consider now the widely-used, one-dimensional harmonic trap potential,
$-\frac{1}{2}\frac{\partial^2}{\partial x^2}+\frac{1}{2}x^2$,
and use the contact interaction $\hat V(x-x')=\lambda_0\delta(x-x')$, 
see \cite{review_Legget,Stringari_book} and references therein. 
We would like to examine here the performance of the CCSD approach.
It should be reemphasized that the CCSD wavefunction contains only a single parameter $c_{22}$.
To solve for this parameter Eq.~(\ref{eq50}) can be used which reduces to the simple quadratic equation
\beq\label{eq57}
 c_{22}^2 \alpha \left(N^2-7N+9\right) + c_{22}\left[(\mu_2-\mu_1)+
\alpha(N-3)+\beta\right] + \alpha/4 = 0 
\eeq
where 
\beqn
 & & \alpha = \lambda_0 \int \left|\varphi_1(x)\right|^2\left|\varphi_2(x)\right|^2 dx \nonumber \\
 & & \beta = \frac{\lambda_0}{2} 
  \left( \int \left|\varphi_1(x)\right|^4 dx + \int \left|\varphi_2(x)\right|^4 dx \right). \nonumber \
\eeqn
Note that $\mu_1$ here is the usual chemical potential of the GP equation.
The ground state energy of the CCSD approach reads
\beq\label{eq58}
 E_0({\mathrm CCSD}) = E_{\mathrm GP} + N(N-1)\alpha c_{22}.
\eeq
Here, $E_{\mathrm GP}=\left<\phi_0\left|H\right|\phi_0\right>$ is the usual ground state GP energy
\beq\label{eq59}
  E_{\mathrm GP} = N \left[h_{11} + \frac{N-1}{2} \lambda_0 \int \left|\varphi_1(x)\right|^4 dx\right].
\eeq

For completeness we would like to compare our CCSD results with those of the CISD. 
The latter wavefunction also contains only one parameter $d_{22}$ (see chapter II)
and the expression for the energy $E_0({\mathrm CISD})$ is identical to that in
Eq.~(\ref{eq58}) if we replace $c_{22}$ by $d_{22}$.
The CISD wavefunction is, however, a superposition of only the two
configurations $\left|N,0\right>$ and $\left|N-2,2\right>$.
The value of the parameter can be simply obtained by diagonalizing the Hamiltonian
$H$ in the space of these two configurations.
This leads to the following quadratic equation
\beq\label{eq60}
 d_{22}^2 \alpha N(N-1) -
 2d_{22} \left[(\mu_2-\mu_1)+\alpha(N-3)+\beta\right] -
 \alpha/2 = 0
\eeq
for the configuration interaction parameter.

The results of our numerical example are summarized in Figs.~1-3.
In Fig.~1 we test the performance of CCSD method in terms of the correlation energy.
The correlation energy is defined as the difference between 
the Gross-Pitaevskii energy, $E_{\mathrm GP}$, and the exact energy, $E_0({\mathrm exact})$. 
The latter is obtained in our model by diagonalizing the many-body 
Hamiltonian within the full configuration-interaction space 
spanned by $\left|m,m'\right>, m=0,...,N, m'=N-m$.
We first calculate the CCSD energy $E_0({\mathrm CCSD})$ using 
Eqs.~(\ref{eq33}-\ref{eq35},\ref{eq57}-\ref{eq59}).
How much of the exact correlation energy $E_{\mathrm GP}-E_0({\mathrm exact})$
is captured by CCSD is given in percent by 
$\%E_{\mathrm correlation}= 100 
\cdot \frac{E_{\mathrm GP}-E_0({\mathrm CCSD})}{E_{\mathrm GP}-E_0({\mathrm exact})}$.
We have calculated $\%E_{\mathrm correlation}$ for $N=100, 1000$ and $10000$
for several values of the interaction strength $\lambda_0$.
The results are plotted in Fig.~1 versus the couping constant $\lambda=\lambda_0(N-1)$,
which is the only interaction parameter entering the GP energy, see Eq.~(\ref{eq59}).
For comparison, the corresponding values obtained by the CISD method
were calculated as well. 
We remind that both methods contain one parameter only, $c_{22}$ and $d_{22}$ respectively.
It is seen that the CCSD is remarkably successful in obtaining
the correlation energy, with absolute error of less than $4\%$ 
up to a huge coupling constant, $\lambda=2\cdot10^4$.
The quality of the CISD, on the other hand, starts to deteriorate
already from $\lambda=1$ on and saturates at about $50\%$, see Fig.~1.
Another result observed in Fig.~1 is that
the performance of CCSD in terms of $\%E_{\mathrm correlation}$ 
improves with increasing $N$, whereas that of CISD worsens.

Next, let us examine the many-body wavefunction obtained by the CCSD method and compare it to the exact one.
For this, we first normalize the CCSD wavefunction (\ref{eq52}) and express it as
$\sum_{m=0}^{N/2} C_{2m} \left|N-2m,2m\right>$.
In Fig.~2 the absolute value of the $C_{2m}$ coefficients
(they alternate in sign because $c_{22}$ is negative) 
for $N=10000$ bosons and $\lambda=100$ are plotted. 
Although the coupling constant is large,
it is remarkable that the CCSD $C_{2m}$ coefficients  
almost perfectly match the exact coefficients 
and it is difficult to distinguish between the red and black curves of Fig.~2.
Another property of the many-body wavefunction when $\lambda$ is growing is
that the tail of the coefficients $C_{2m}$ is extending further,
showing that more and more excited configurations contribute to the many-body wavefunction. 
For comparison, the two coefficients of the CISD are also shown,
which deviate much from the exact solution, see Fig.~2.

Finally, we examine the capability of the CCSD method to reproduce the exact ground-state 
depletion, i.e., the average number $\left<n_2\right>$ of bosons
occupying the orbital $\varphi_2$.
As mentioned above, $\left<n_2\right>$ and $\left<n_1\right>=N-\left<n_2\right>$
are the eigenvalues of the reduced one-body density and hence
are a very sensitive tool for the quality of the CCSD many-body wavefunction.  

We have calculated $\left<n_2\right>$ for $N=100, 1000$ and $10000$
for several values of the interaction strength $\lambda_0$ 
up to the huge value of $\lambda=2\cdot10^4$.
The results are plotted in Fig.~3 together with
the exact ones and the corresponding values obtained with CISD.
It is seen that the CCSD is extremely successful in obtaining
the depletion $\left<n_2\right>$ up to a large coupling constant, $\lambda=10^2$.
From about this value on, the quality of the CCSD wavefunction in terms of
$\left<n_2\right>$ depends on the specific number $N$ of bosons.
For $N=100$ it is very good for all values of $\lambda$.
For $N=10000$ at the extreme value $\lambda=2\cdot10^4$
it predicts 2.5 as much depletion as the exact many-body wavefunction gives,
namely almost 20 bosons instead of 8 out of 10000 bosons.
We remind that all these results are obtained with a single parameter $c_{22}$.
The deviations of $\left<n_2\right>$ for large $N$ and larger interaction strength 
$\lambda_0$ is related to the tail of the $C_{2m}$ distribution.
As $N$ and $\lambda_0$ increase, 
there are more and more non-negligible
CCSD coefficients which start to deviate from the exact ones.
While this does not lead to an error of more than $3\%$ percent in the correlation energy,
see Fig.~1, it does influence the more sensitive measure of exactness of the wavefunction, $\left<n_2\right>$.
For comparison, we also computed the corresponding $\left<n_2\right>$ values with CISD
and plotted the results in Fig.~3.
We obtained that the values of $\left<n_2\right>$ for all $N$ saturates at about $0.18$
with increasing $\lambda$, which is more than an order of magnitude smaller than the exact and CCSD results.
This near independence of $\left<n_2\right>$ in CISD from the number of bosons $N$ 
is a manifestation of the minimal correlations
embedded in the CISD wavefunction, in contrast to the CCSD wavefunction.

Summarizing the results depicted in Figs.~1-3, we see that the CCSD 
for bosons performs remarkably well even for large interaction strengths.
Utilization of the ground configuration in Eq.~(\ref{eq7})
is an appropriate choice for the coupled-cluster expansion
at least for this example
(see also the discussion below).

\section{Summary and conclusions}

The standard configuration-interaction approach rapidly becomes impractical in the many-body problem.
When the interaction between $N$ bosons is substantial and/or many of them are present,
the number of configurations necessary to correctly describe the correlated wavefunction
quickly increases beyond computational reach.
In searching for more efficient approaches which are amenable to systematic approximations
(truncations), we have developed in this paper a coupled-cluster theory
for systems of bosons in external traps.

In the coupled-cluster approach the exact wavefunction is obtained by 
applying an exponential operator $\exp\{T\}$ to the ground configuration $\left|\phi_0\right>$.
The ground configuration $\left|\phi_0\right>$ depends, of course, on the particle statistics. 
While for fermions it is a determinant with $M=N$ different orbitals, 
the situation for bosons is more intricate. 
Since there is no limitation on the number of bosons occupying a certain orbital,
there are ample legitimate choices for the ground permanent of $N$ interacting bosons 
over $M$ available orbitals. 
The most natural choice for non-interacting or weakly-interacting bosons
is, of course, to let all bosons reside in the orbital lowest in energy $\varphi_1$,
$\left|\phi_0\right>=\frac{1}{\sqrt{N!}} (b_1^\dag)^N\left|0\right>$,
which is our main choice for the coupled-cluster theory presented here. 

Because of the simple structure of $\left|\phi_0\right>$, 
the appearance of excitation operators $T=\sum_{n=1}^N T_n$ for bosons is much simpler than for fermions.
When the simplest truncation $T=T_1$ is chosen, namely CCS,
the effect of $\exp\{T_1\}$ on $\left|\phi_0\right>$ 
is to transform $\varphi_1$ to another orbital $\tilde \varphi_1$.
$\exp\{T_1\}$ optimizes this orbital by mixing the $M$ available orbitals.
This reminds the situation encountered for fermions, 
where the operation of the fermionic $T_1$ transforms the ground determinant into
another determinant (Thouless theorem \cite{Thouless}).

In a substantial part of this work we addressed the issue of
size consistency for bosons and enquired whether truncated coupled-cluster expansions are size consistent.
It turns out that the answer to this question depends on the choice of  
ground configuration (permanent). 
Considering $R$ non-interacting replica of the $N$-boson system, 
it has been found that truncated coupled-cluster expansions {\it are not} size consistent
with the simplest choice for the $R$-replica ground permanent,
$\left|\phi_0\right> = \frac{1}{\sqrt{(NR)!}}\left(b_1^\dag\right)^{NR} \left|0\right>$,
already for the mean-field energy $\left<\phi_0\left|H\right|\phi_0\right>$.
This is a surprising result when compared to the case of fermions.
Fortunately, this violation of size consistency, at least for the mean-field energy, 
leads to negligible errors for large individual systems ($N\gg 1)$.
Can size consistency in bosonic systems be fully restored, 
perhaps with another choice of the $R$-replica ground permanent?
Yes, it has been straightforwardly shown that truncated coupled-cluster expansions are 
size consistent with the ground permanent 
$\left|\phi_0\right> = \frac{1}{(N!)^{R/2}} \prod_{k=1}^R \left(b_{1_k}^\dag\right)^N \left|0\right>$,
which ``distinguishes'' between the $R$ replica; also see discussion below.

Next, we moved to derive working equations of the coupled-cluster approach
for the natural ground configuration $\left|\phi_0\right>=\frac{1}{\sqrt{N!}}(b_1^\dag)^N\left|0\right>$.
First, it has been shown that the exact correlation energy depends on two kinds of coefficients only: 
$\{c_k\}$ and $\{c_{kl}\}$ of the single and double excitation operators $T_1$ and $T_2$.
For a given truncation of the coupled-cluster exponential operator $\exp\{T\}$,
it is possible in principle to calculate the correlation energy.
Here, for the specific truncation of $T=T_1+T_2$, i.e., CCSD, 
general working equations for $\{c_k\}$, $\{c_{kl}\}$ have been explicitly derived.
 
Finally, we tested the performance of the CCSD in a model where an exact solution can be computed.
We employed an harmonic trap and restricted the orbital space to two orbitals of different spatial symmetry.
The exact solution is obtained, of course, by diagonalizing the
many-body Hamiltonian within the full configuration-interaction space 
spanned by $\left|m,m'\right>, m=0,...,N, m'=N-m$.
In contrast, the CCSD approach requires here one parameter only, $c_{22}$,
which is a solution of a simple algebraic equation of the second degree, see Eq.~(\ref{eq57}).
The performance of the CCSD approach for $N=100, 1000$ and $10000$ interacting bosons 
was tested in terms of three criteria: correlation energy, many-body wavefunction and ground-state depletion.
It was found that the CCSD is remarkably successful in obtaining
the correlation energy, with absolute error of less than $4\%$ 
up to a huge coupling constant, $\lambda=2\cdot10^4$, see Fig.~1.
The quality of the CCSD many-body wavefunction and its ability to
accurately describe ground-state depletion were found to be   
remarkably good for all boson numbers and coupling constants as 
large as $\lambda=10^2$, see Figs.~2-3.
For comparison, we examined the performance of CISD,
which similarly depends on one parameter only, $d_{22}$.
CISD was found to be substantially poorer in comparison to CCSD.
For instance, it accounts for about $50\%$ of the correlation energy only.

The coupled-cluster theory for bosons presented in this work,
as certainly supported by the numerical example, 
is a promising approach to be further developed in the many-boson problem.
The expressions of the bosonic coupled-cluster theory are 
much simpler than those for fermions since, generally, 
the ground configuration (permanent) employs one orbital only.
Consequently, we can treat a very large number of bosons
with coupled-cluster expansions and employ more virtual (non-occupied) orbitals
than the fermionic coupled-cluster can.
These qualities open the way to study few- to many-boson systems
up to a substantial interaction where several orbitals
are needed to describe the reality.

The issue of size consistency, as extensively discussed above, is 
delicate for bosons, and depends on the choice of the ground configuration.
It relates to the following practical point:
what is a suitable choice of the ground permanent 
when a coupled-cluster expansion is to be employed
with a specific physical system?
We can say that, for bosons in a single-well trap
an useful choice is the simplest permanent where all bosons
reside in the same orbital, which is the standard mean-field, Gross-Pitaevskii orbital.
However, if we wish to usefully apply coupled-cluster expansions 
to a bosonic system undergoing spatial fragmentation or 
superfluid to Mott-insulator transitions, situations that occur in double- and 
multi-well traps, we have to be more careful with the choice of ground configuration,
and depart from the simplest permanent constructed from the Gross-Pitaevskii mean-field orbital.
Recently, a more general mean-field theory has been introduced,
allowing for bosons to reside in several orbitals \cite{LA_OAL_PLA}.
We anticipate that in combination with coupled-cluster expansions they 
can be useful for studying many bosons in double- and multi-well traps.   

\acknowledgments
The authors wish to thank Nayana Vaval for comments.

\begin{appendix}
\section{} 

The transformed two-body operator $\dot V$ of the transformed Hamiltonian
$\dot H = \dot H_0 + \dot V$ consists of the nine terms listed below.
$$
 \dot V = \sum_{p=1}^{9} \dot V(p)
$$
\beqn
 & & \dot V(1) = \frac{1}{2} V_{1111} \dot b_1^\dag \dot b_1^\dag b_1 b_1 \nonumber \\
 & & \dot V(2) = \sum_{k=2}^M V_{111k} \dot b_1^\dag \dot b_1^\dag b_1 \dot b_k \nonumber \\
 & & \dot V(3) = \sum_{k=2}^M V_{k111} b_k^\dag \dot b_1^\dag b_1 b_1 \nonumber \\
 & & \dot V(4) = \frac{1}{2} \sum_{k,l=2}^M V_{11kl} \dot b_1^\dag \dot b_1^\dag \dot b_k \dot b_l \nonumber \\
 & & \dot V(5) = \frac{1}{2} \sum_{k,l=2}^M V_{kl11} b_k^\dag b_l^\dag b_1 b_1 \nonumber \\
 & & \dot V(6) = \sum_{k,l=2}^M \left(V_{1k1l}+V_{1kl1}\right) \dot b_1^\dag b_k^\dag b_1 \dot b_l \nonumber \\
 & & \dot V(7) = \sum_{j,k,l=2}^M V_{1jkl} \dot b_1^\dag b_j^\dag \dot b_k \dot b_l \nonumber \\
 & & \dot V(8) = \sum_{j,k,l=2}^M V_{jkl1} b_j^\dag b_k^\dag \dot b_l b_1 \nonumber \\
 & & \dot V(9) = \frac{1}{2} \sum_{i,j,k,l=2}^M V_{ijkl} b_i^\dag b_j^\dag \dot b_k \dot b_l \nonumber \
\eeqn

\end{appendix}

\begin{figure}[ht] 
\includegraphics[width=11cm,angle=-0]{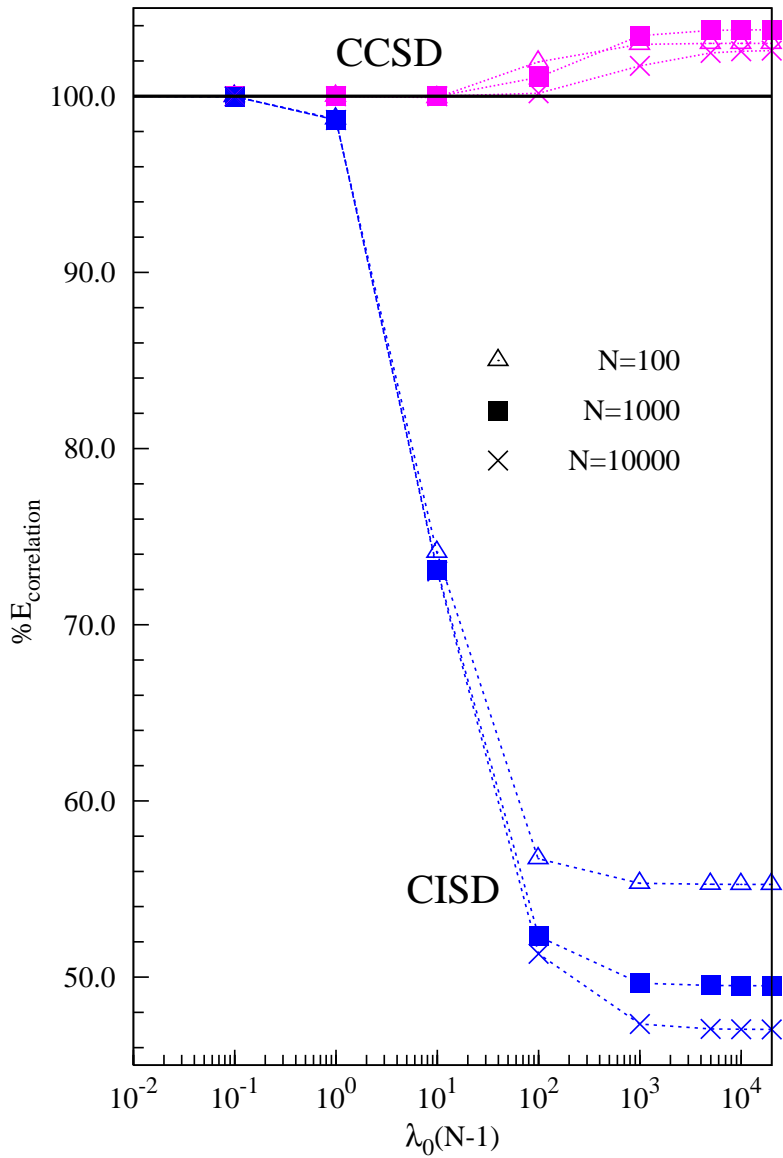}
\vglue 0.5 truecm
\caption [kdv] {(Color online) Performance of CCSD method: correlation energy.
Shown is the percent of correlation energy, denoted by $\%E_{\mathrm correlation}$,
obtained by the CCSD for $N=100,1000,10000$ bosons
and several values of interaction strength $\lambda_0$.
The correlation energy is defined as the difference between 
the Gross-Pitaevskii energy $E_{\mathrm GP}$ and the exact energy.
The exact energy is obtained in our model by diagonalizing the many-body 
Hamiltonian within the full configuration-interaction space. 
For comparison, the corresponding values obtained by CISD are also plotted. 
} 
\end{figure}

\begin{figure}[ht] 
\includegraphics[width=12cm,angle=-0]{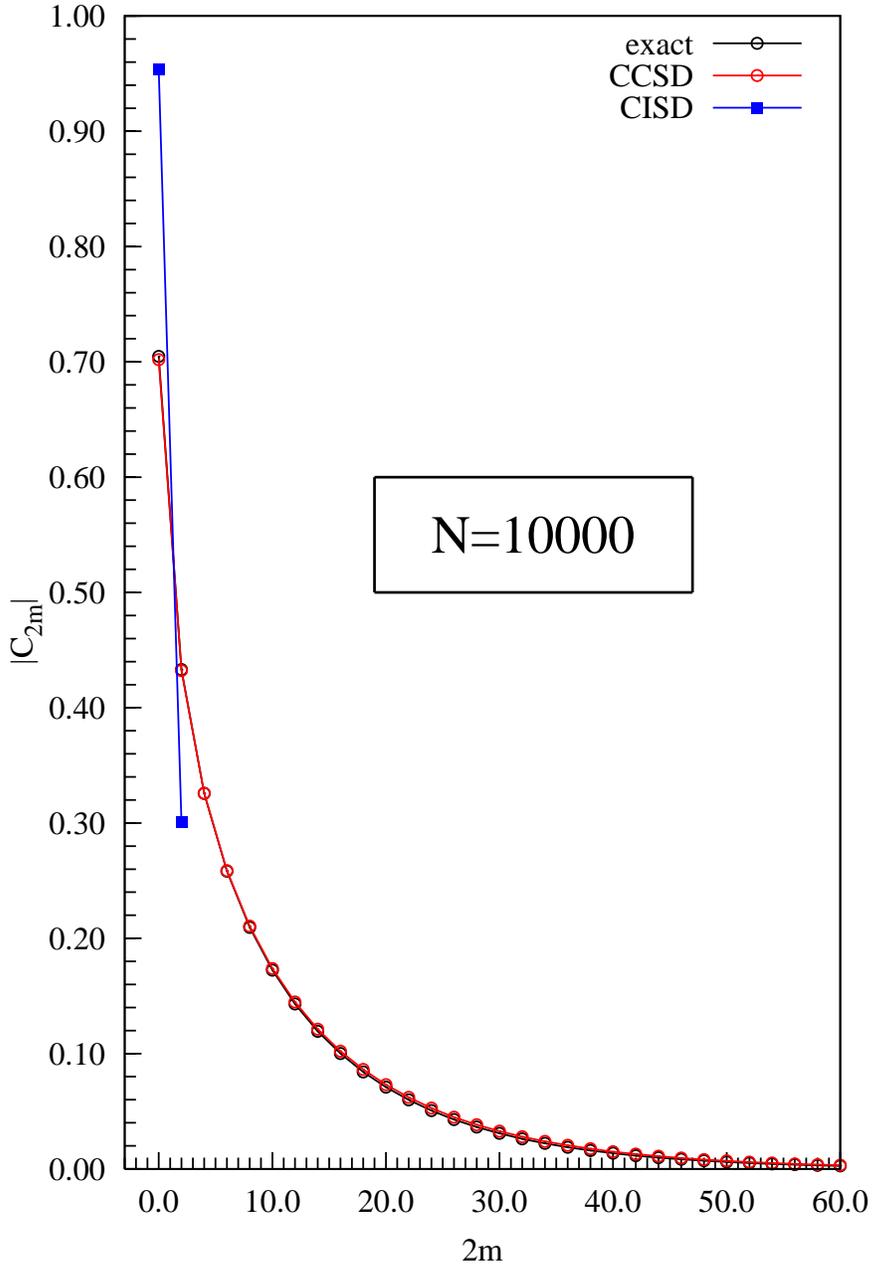}
\vglue 0.5 truecm
\caption [kdv] {(Color online) Performance of CCSD method: many-body wavefunction. 
Shown are the coefficients, $C_{2m}$, of the normalized many-body wavefunction,
$\sum_{m=0}^{N/2} C_{2m} \left|N-2m,2m\right>$, 
for $N=10000$ bosons and $\lambda=\lambda_0(N-1)=100$ 
obtained by the CCSD method, see Eqs.~(\ref{eq52},\ref{eq53}).
Although the coupling constant is large,
it is remarkable that the CCSD $C_{2m}$ coefficients  
almost perfectly match the exact coefficients, 
namely the red curve ``sits'' atop the black curve.
For comparison, the two coefficients of the CISD are also shown,
which deviate much from the exact solution.}
\end{figure}	

\begin{figure}[ht] 
\includegraphics[width=11cm,angle=-0]{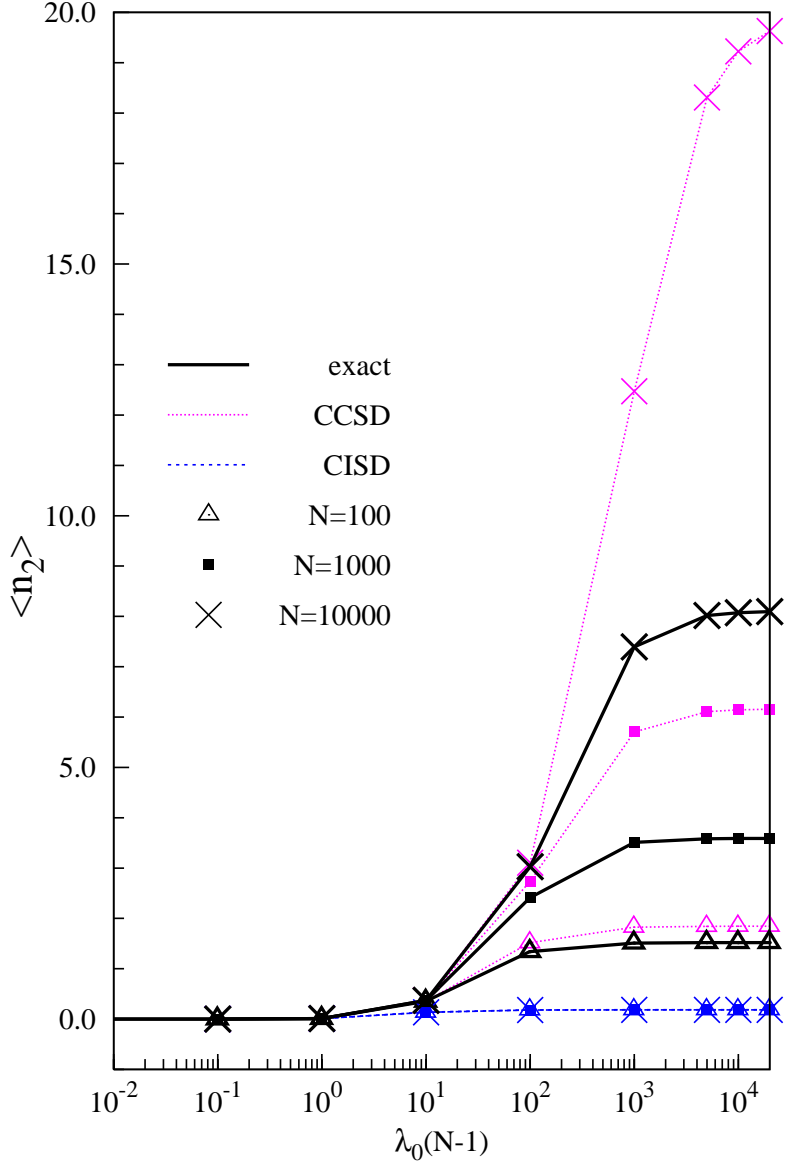}
\vglue 0.5 truecm
\caption [kdv] {(Color online) Performance of CCSD method: ground-state depletion.
Shown is the average number of bosons in orbital $\varphi_2$, 
$\left<n_2\right>$, for $N=100,1000,10000$ bosons
and several values of interaction strength.
Because of the different spatial symmetry of $\varphi_1$ and $\varphi_2$, these orbitals are 
 the eigenfunctions of the reduced one-particle density matrix (natural orbitals)
and the respective eigenvalues are $\left<n_1\right>$ and $\left<n_2\right>$
with $\left<n_1\right>+\left<n_2\right>=N$.  
It is seen that the CCSD is extremely successful in obtaining
the depletion $\left<n_2\right>$, which is a very sensitive measure of the exactness of the
many-body wavefunction, for all $N$ up to a large coupling constant, $\lambda=10^2$.
The CISD results are also shown for comparison.
See text for more details.}
\end{figure}

\ed